\documentstyle[amsfonts,prd,aps,epsf]{revtex}
\begin{document}
\title{Dynamical $N$-body Equlibrium in Circular Dilaton Gravity}
\author{R. Kerner$^{(a,b)}$ and R. B. Mann$^{(a)}$ }
\address{$^{(a)}$Dept. of Physics, University of Waterloo, Waterloo, Ontario%
\\
N2L 3G1, Canada}
\address{$^{(b)}$Dept. of Physics, University of Alberta, Edmonton, Alberta\\
V6T 1W5, Canada}
\maketitle

\begin{abstract}
We obtain a new exact equilibrium solution to the $N$-body problem in a
one-dimensional relativistic self-gravitating system. \ It corresponds to an
expanding/contracting spacetime of a circle with $N$ bodies at equal proper
separations from one another around the circle. \ \ Our methods are
straightforwardly generalizable to other dilatonic theories of gravity, and
provide a new class of solutions to further the study of (relativistic)
one-dimensional self-gravitating systems.
\end{abstract}

%\thispagestyle{empty}
%======================================%
%<<<<<<<<<<<< TITLE PAGE >>>>>>>>>>>>>>%
%======================================%
%\thispagestyle{empty}
%{\baselineskip0pt
%\leftline{\large\baselineskip16pt\sl\vbox to0pt{\hbox{RESCEU}
%               \hbox{The University of Tokyo}\vss}}
%\rightline{\large\today}
%}
%\vskip20mm
%\twocolumn[\hsize\textwidth\columnwidth\hsize\csname
%@twocolumnfalse\endcsname

\bigskip

%\vfill
%======================================%
%<<<<<<<<<<<< SECTION I  >>>>>>>>>>>>>>%
%======================================%
%\baselineskip25pt
Circular gravity is a general relativistic gravitational system in $(1+1)$
dimensions in which the topology of space is equivalent to that of a circle $%
S^{1}$. \ Its counterpart, lineal gravity (the same system with the $S^{1}$
topology replaced by ${\Bbb R}^{1}$), has been the subject of much
investigation over the past twenty years, motivated by the desire to
understand black holes \cite{r3}, spacetime structure \cite{browndan,Strobl}%
, canonical quantum gravity \cite{Gabor}, string physics \cite{Witten},
information loss \cite{CGHS} and the $N$-body problem \cite{OR,pchak}. This
latter problem involves consideration of a collection of $\ N$ particles
interacting through their own mutual gravitational attraction, along with
other specified forces. \ Apart from functioning as prototypes for higher
dimensional gravity, in lineal gravity such self-gravitating systems also
approximate the behaviour of some physical systems in 3 spatial dimensions,
such as stellar dynamics orthogonal to a galactic plane, the collisions of
flat parallel domain walls, and the dynamics of cosmic strings. The ergodic
and equipartition properties of non-relativistic one-dimensional
self-gravitating systems are a current subject of research \cite{yawn}, and
recently such systems were shown to exhibit fractal behaviour \cite{fractal}%
. \ In the past few years these systems have been extended to relativistic
lineal gravity. The 2-body problem has been solved exactly \cite{2bd}, and
has been extended to include both cosmological expansion \cite%
{2bdcossh,2bdcoslo}\ and electromagnetic interactions \cite{2bdchglo}. \ \
The $N$-body extension of these systems, while considerably more
complicated, is nevertheless amenable to analysis \cite{pchak} and includes
the non-relativistic self-gravitating systems \cite{yawn,Rybicki} as a
limiting case.

Comparitively little is known about the circular counterparts of these
systems; they introduce qualitatively new features that are absent in the
non-compact lineal case. For example, in the absence of other fields there
are no static non-relativistic equilibrium solutions for $N$ particles on a
circle. This is because the non-relativistic potential grows linearly with
increasing distance from the source, and a circular topology admits no
solutions to the matching conditions for any $N\geq 1$.

We present here an exact equilibrium solution to the $N$-body problem in
circular gravity. Our solution corresponds to $N$ equal mass particles at
equidistant proper separations from one another in a spacetime that
expands/contracts in response to these sources. \ The configuration we find
therefore corresponds to a dynamic, albeit unstable, equilibrium. To our
knowledge these are the first $N$-body dynamic equilibrium solutions in any
relativistic theory of gravity. While our solutions depend upon the choice
of dilaton potential, our approach is straightforwardly generalizable to any
dilatonic theory of gravity, and provides a new class of solutions to
further the study of (relativistic) one-dimensional self-gravitating systems.

The action integral for a generic dilaton theory of gravity \cite%
{Gabor,banksmann} minimally coupled to $N$ point masses is 
\begin{eqnarray}
I &=&\int d^{2}x\left[ \frac{1}{2\kappa }\sqrt{-g}g^{\mu \nu }\left\{ \Psi
R_{\mu \nu }+\frac{1}{2}\nabla _{\mu }\Psi \nabla _{\nu }\Psi +g_{\mu \nu
}V\left( \Psi \right) \right\} \right.  \nonumber \\
&&\qquad \left. +\sum_{a}\int d\tau _{a}\left\{ -m_{a}\left( -g_{\mu \nu }(x)%
\frac{dz_{a}^{\mu }}{d\tau _{a}}\frac{dz_{a}^{\nu }}{d\tau _{a}}\right)
^{1/2}\right\} \delta ^{2}(x-z_{a}(\tau _{a}))\right] \;,  \label{act0}
\end{eqnarray}%
where $\Psi $ is the dilaton field, $g_{\mu \nu }$ is the metric with
determinant $g$ and covariant derivative $\nabla _{\mu }$ , $R$ is the Ricci
scalar, and $\tau _{a}$ is the proper time of $a$-th particle, with $\kappa
=8\pi G/c^{4}$. For circular gravity the range of the spatial coordinate $x$
is $-L\leq x\leq L$, and the circular topology implies that all fields must
be smooth (or at least $C^{1}$) functions of $x$ with period $2L$, which
implies 
\begin{equation}
f(L)=f(-L)\qquad \text{and}\qquad f^{\prime }(L)=f^{\prime }(-L)=0\;\;.
\label{smooth}
\end{equation}%
for all functions. Henceforth we set $V\left( \Psi \right) =\Lambda /2$
throughout, where $\Lambda $ is constant.

The action (\ref{act0}) describes a generally covariant self-gravitating
system (without collisional terms, so that the bodies pass through each
other); its field equations are \noindent 
\begin{equation}
R-\Lambda =\kappa T_{\;\;\mu }^{P\mu }\quad \frac{d}{d\tau _{a}}\left\{ 
\frac{dz_{a}^{\nu }}{d\tau _{a}}\right\} +\Gamma _{\alpha \beta }^{\nu
}(z_{a})\frac{dz_{a}^{\alpha }}{d\tau _{a}}\frac{dz_{a}^{\beta }}{d\tau _{a}}%
=0  \label{RTgeo}
\end{equation}%
\begin{equation}
\frac{1}{2}\nabla _{\mu }\Psi \nabla _{\nu }\Psi -g_{\mu \nu }\left( \frac{1%
}{4}\nabla ^{\lambda }\Psi \nabla _{\lambda }\Psi -\nabla ^{2}\Psi \right)
-\nabla _{\mu }\nabla _{\nu }\Psi =\kappa T_{\mu \nu }^{P}+\frac{\Lambda }{2}%
g_{\mu \nu }  \label{e4}
\end{equation}%
where the stress-energy due to the point masses is 
\begin{equation}
T_{\mu \nu }^{P}=\sum_{a=1}^{N}m_{a}\int d\tau _{a}\frac{1}{\sqrt{-g}}g_{\mu
\sigma }g_{\nu \rho }\frac{dz_{a}^{\sigma }}{d\tau _{a}}\frac{dz_{a}^{\rho }%
}{d\tau _{a}}\delta ^{2}(x-z_{a}(\tau _{a}))  \nonumber
\end{equation}%
and is conserved. The preceding system generalizes Jackiw-Teitelboim lineal
gravity \cite{JT}, which equates scalar curvature to a (cosmological)
constant $\Lambda $, reducing to it if the stress-energy vanishes. The
specific choice of dilaton potential ensures that the evolution of the
dilaton does not modify the aformentioned reciprocal gravity/matter
dynamics. \ The system (\ref{RTgeo}) is a closed system of $N+1$ that
describes the evolution of the single metric degree of freedom and the $N$
degrees of freedom of the point masses. The evolution of the point-masses
governs the evolution of the dilaton via (\ref{e4}), whose
divergencelessness is consistent with the conservation of $T_{\mu \nu }^{P}$%
, yielding only one independent equation to determine the single degree of
freedom of the dilaton.

Making use of the previously formulated general framework for canonical
circular dilaton gravity \cite{circle}, we write the metric as $%
ds^{2}=-\left( N_{0}dt\right) ^{2}+\gamma \left( dx+\frac{N_{1}}{\gamma }%
dt\right) ^{2}$. The extrinsic curvature $K$ is then%
\begin{equation}
K=(2N_{0}\gamma )^{-1}(2\partial _{1}N_{1}-\gamma ^{-1}N_{1}\partial
_{1}\gamma -\partial _{0}\gamma )=\sqrt{\gamma }\kappa (\pi -\Pi /\gamma )
\label{Kext}
\end{equation}%
where the conjugate momenta to $\gamma $ and $\Psi $ are respectively
denoted $\pi $ and $\Pi $. \ We can take the extrinsic curvature $K$ to be a
time coordinate $\tau \left( t\right) $ of the circle, thereby allowing the
elimination of $\pi $ from all field equations and the identification of the
Hamiltonian from the action (\ref{act0}) after canonical reduction \cite%
{circle}: 
\begin{equation}
I=\int d^{2}x\left\{ \sum_{a}p_{a}\dot{z}_{a}\delta (x-z_{a})+\Pi \frac{%
\partial }{\partial t}\left( \Psi +\ln \gamma \right) -{\cal H}\right\} \;,
\label{act3}
\end{equation}%
where $H=\int dx{\cal H=}\frac{2\dot{\tau}}{\kappa }\int dx\sqrt{\gamma }$,
which is the circumference functional of the circle when $\dot{\tau}$ is
constant. \ The dynamics is that of a time-dependent system, with the
time-dependence corresponding to the time-varying circumference of the
circle of constant mean extrinsic curvature. As with general relativity on
spatially compact manifolds in $(2+1)$ dimensions \cite{Moncreif}, \ the\
spatial metric can be chosen so that $\gamma =\gamma (t)$ , i.e. \ $\gamma
^{\prime }=0\;$, since any metric on a circle is globally conformal to a
flat metric. Choosing a time parametrization so that $\dot{\tau}\sqrt{\gamma 
}$ is constant, the Hamiltonian will be time-independent.

We find after some manipulation that the canonical form of the field
equations (\ref{RTgeo},\ref{e4}) is 
\begin{eqnarray}
2\Pi ^{\prime }-\Pi \Psi ^{\prime }=0; &&  \label{fin9} \\
\Psi ^{\prime \prime }-\frac{1}{4}(\Psi ^{\prime })^{2}-(\kappa \Pi
)^{2}+\gamma \left( \tau ^{2}-\Lambda /2\right) +\kappa \sum_{a}\sqrt{\gamma 
}\;m\;\delta (x-z_{a}(x^{0}))=0 &&  \label{fin10} \\
N_{0}^{\prime \prime }-\dot{\tau}\gamma -N_{0}\left\{ (\tau ^{2}-\Lambda
/2)\gamma +\kappa /2\sum_{a}\sqrt{\gamma }\;m\;\delta
(x-z_{a}(x^{0}))\right\} =0 &&  \label{fin11} \\
N_{1}^{\prime }-\dot{\gamma}/2+\gamma \tau N_{0}=0 &&  \label{fin12} \\
\dot{\Pi}+\partial _{1}(-\frac{1}{\gamma }N_{1}\Pi +\frac{1}{2\kappa \sqrt{%
\gamma }}N_{0}\Psi ^{\prime }+\frac{1}{\kappa \sqrt{\gamma }}N_{0}^{\prime
})=0 &&  \label{fin13} \\
\dot{\Psi}+2N_{0}\left( \kappa \frac{\Pi }{\sqrt{\gamma }}+\tau \right)
-N_{1}(\frac{1}{\gamma }\Psi ^{\prime })=0 &&  \label{fin14}
\end{eqnarray}%
where equilibrium solutions are characterized by $\dot{z_{a}}=0=p_{a}$,
corresponding to a situation in which all particles are motionless at
various points around the circle. A calcuation of the geodesic equations in (%
\ref{RTgeo}) show that this implies 
\begin{equation}
\left. \frac{\partial N_{0}}{\partial x}\right| _{x=z_{a}}=0\text{ \ \ \ \ \
\ \ \ \ \ }N_{1}(z_{a})=0\;\;.  \label{eqzeqs}
\end{equation}

Remarkably, this set of equations admits an exact solution for $N$ bodies of
equal mass $m$ at equal time-varying proper separations. We set $c^{2}\equiv
\gamma \left( \tau ^{2}-\Lambda /2\right) =c_{+}^{2}>0$, and $r=\left|
z_{a}-z_{a+1}\right| =\frac{2L}{N}$ , where $a=1,2,\ldots ,N$. \ The system
is cyclically symmetric and so we can choose the origin to be halfway
between any two particles in the $N=$even case or on a particle in the $N=$%
odd case. In seeking solutions to eqs. (\ref{fin9}--\ref{fin14}) we must
ensure that the solutions for $\Psi $ and $N_{0}$ are appropriately matched
at the locations $z_{a}$ of each particle and the identification point $%
\left| x\right| =L$; they must be continuous there but not differentiable at
the locations of the particles; for example $\lim_{\epsilon \rightarrow 0}%
\left[ \Psi ^{\prime }(z_{a}+\epsilon )-\Psi ^{\prime }(z_{a}-\epsilon )%
\right] =\kappa \sqrt{\gamma }m$.

With this in mind, eq. (\ref{fin9}) is straightforwardly found to yield $\Pi
=\Pi _{0}(t)\exp {\Psi /2}$, where $\Pi _{0}(t)$ is an arbitrary function. \
Upon substitution of this result into eq. (\ref{fin10}), and employing the
conditions (\ref{eqzeqs}), the next three equations are solved by
straightforward integration and matching across the particles, furnishing a
solution \ for ${\Psi },N_{0},$ and $N_{1}$ in terms of \ $\Pi _{0}\left(
t\right) $ and the metric function $\gamma \left( t\right) $. \ Imposing (%
\ref{eqzeqs}) and the periodicity condition $N_{1}(L)=N_{1}(-L)$ then yields
a differential equation for $\gamma $ whose solution is 
\begin{equation}
\sqrt{\gamma }=\frac{N}{L\sqrt{\tau ^{2}-\Lambda /2}}\text{arctanh}\left( 
\frac{\xi \sqrt{\frac{\kappa ^{2}M^{2}}{16N^{2}}+(\xi ^{2}-1)(\tau
^{2}-\Lambda /2)}-\frac{\kappa M}{4N}}{\frac{\kappa ^{2}M^{2}}{16N^{2}}+\xi
^{2}(\tau ^{2}-\Lambda /2)}\sqrt{\tau ^{2}-\Lambda /2}\right)
\label{gamnbody}
\end{equation}%
where $M=mN$ is the total mass of the system. The function $\Pi _{0}\left(
t\right) $ is then found from integration of (\ref{fin14}). Satisfaction of
equation (\ref{fin13}) provides a non-trivial consistency check on the full
result, which is

\begin{eqnarray}
\Psi &=&-2\ln \left( \beta -\frac{\sinh (\frac{c_{+}L}{N})}{\sinh (c_{+}L)}%
\sum_{a=1}^{n}\cosh (c_{+}(|x-z_{a}|-L))\right) -2\ln (\frac{\kappa \Pi
_{0}\left( t\right) }{c_{+}\sqrt{\beta ^{2}-1}})  \label{psisum} \\
N_{0} &=&\frac{\dot{\tau}\gamma }{c_{+}^{2}\beta }\left( \frac{\sinh (\frac{%
c_{+}L}{N})}{\sinh (c_{+}L)}\sum_{a=1}^{n}\cosh (c_{+}(|x-z_{a}|-L)))-\beta
\right)  \label{lapsesum} \\
N_{1} &=&\gamma \frac{\dot{c_{+}}}{c_{+}}x-\frac{\gamma ^{2}\tau \dot{\tau}}{%
c_{+}^{3}\beta }\frac{\sinh (\frac{c_{+}L}{N})}{\sinh (c_{+}L)}%
\sum_{a=1}^{n}\epsilon \left( x-z_{a}\right) \left[ \sinh
(c_{+}(|x-z_{a}|-L))+\sinh (c_{+}L)\right]  \label{shiftsum}
\end{eqnarray}%
for the dilaton, lapse and shift functions, where the step function $%
\epsilon \left( x\right) =\frac{\left| x\right| }{x}$, and $\epsilon \left(
0\right) =0$. \ \ Here%
\begin{eqnarray}
\beta &=&\frac{4N}{\kappa M}\sqrt{\frac{\kappa ^{2}M^{2}}{16N^{2}}+(\xi
^{2}-1)(\tau ^{2}-\Lambda /2)}  \label{betanbodysol} \\
\Pi _{0} &=&\sqrt{\gamma }\zeta ^{2}\left( \tau \pm \sqrt{\frac{\kappa
^{2}M^{2}}{16N^{2}(\xi ^{2}-1)}+(\tau ^{2}-\Lambda /2)}\right)
\label{Pi0nbodysol}
\end{eqnarray}%
with $\zeta $ and $\xi $ integration constants. \ The conjugate dilaton
momentum is easily seen to be 
\begin{equation}
\kappa \Pi =\frac{\pm c_{+}\sqrt{\beta ^{2}-1}}{\frac{\sinh (\frac{c_{+}L}{N}%
)}{\sinh (c_{+}L)}\sum_{a=1}^{n}\cosh (c_{+}(|x-z_{a}|-L))-\beta }
\label{Pinbody}
\end{equation}

We can more succinctly write the above using the relation $%
\sum_{a=1}^{n}\cosh (c_{+}(|x-z_{a}|-L))=\cosh (c_{+}f(x))\frac{\sinh
(c_{+}L)}{\sinh (\frac{c_{+}L}{N})}$ where $f(x)$ is the saw-tooth function
that peaks with a value of $L/N$ (i.e. $f(z_{a})=L/N$) at the particle
locations and has zero magnitude half-way between the particles (i.e. $f(%
\frac{z_{a}-z_{a+1}}{2})=0$ ). This relation holds because each function is
periodic, and so the origin can be shifted to be between any pair of
adjacent particles. A simple shift of the origin in the cosh function and a
subsequent manipulation of the sum yields the equivalence.

\bigskip

Another class of solutions exists for $c^{2}=c_{-}^{2}<0$. These are
obtained upon replacing $c_{+}\rightarrow ic_{-}$ in the above results. \
The solution for the metric function $\gamma $\ now becomes%
\begin{equation}
\sqrt{\gamma }=\frac{N}{L\sqrt{\Lambda /2-\tau ^{2}}}\left[ \arctan \left( 
\frac{\xi \sqrt{\frac{\kappa ^{2}M^{2}}{16N^{2}}+(1-\xi ^{2})(\Lambda
/2-\tau ^{2})}-\frac{\kappa M}{4N}}{\frac{\kappa ^{2}M^{2}}{16N^{2}}-\xi
^{2}(\Lambda /2-\tau ^{2})}\sqrt{\Lambda /2-\tau ^{2}}\right) +k\pi \right]
\label{gamnbodyneg}
\end{equation}%
where $k$ is an integer and the integration constant $\left| \xi \right| <%
\sqrt{1+\frac{\left( \kappa M\right) ^{2}}{8\Lambda N^{2}}}$ as a
consequence of the periodicity of $N_{1}(x,t)$. \ There are no solutions for 
$c^{2}=0$.

The solution (\ref{gamnbody}--\ref{shiftsum},\ref{Pinbody}) (and its $%
c^{2}<0 $ counterpart) is a new dynamical equilibrium solution to the $N$%
-body problem in relativistic gravity. \ It corresponds to an
expanding/contracting spacetime of a circle with $N$ bodies at equal
time-varying proper separations from one another around the circle. For $N=1$
the results reduce to the single particle case obtained previously \cite%
{circle}; indeed, the $N$-body solution for the spatial metric is equivalent
to that of the single-particle solution with a rescaling of the parameters $%
L\rightarrow L/N$ and $M\rightarrow M/N$. Note that this rescaling
equivalence does not hold for the remaining metric and dilaton functions. \ 

Analysis of the behaviour of the spatial metric and Hamiltonian thus
proceeds along the same lines as ref.\cite{circle}. For fixed particle mass $%
m$ the spatial metric and Hamiltonian behave analogously to the
single-particle case times a factor of $N$. \ For $\Lambda <0$ the proper
circumference of the circle expands from zero to some maximal size and then
recontracts, whereas for $\Lambda =0$ it undergoes a perturally decelerating
expansion due to the presence of the point masses. \ For $\Lambda >0$ the
cosmological expansion opposes decelerating effects due to the point masses
and a variety of behaviours can ensue, including expansions from zero size
that asymptote to some finite value, evolutions to maximal size that
begin/end at finite circumference and perpetually oscillating universes. \ 

For fixed total mass $M$ the evolution of the spatial metric becomes
independent of $M$ as the number of bodies $N$ becomes large, and the full
solution more closely approaches its vacuum state on a circle with $%
L\rightarrow L/N$. \ 

Generalization of our approach to other dilatonic theories of gravity is
straightforward. An arbitrary $(1+1)$-dimensional dilatonic theory of
gravity minimally coupled to $N$ bodies can, after conformal transformations
and a rescaling of the dilaton, be written in the form (\ref{act0}) \cite%
{Gabor,banksmann} with the replacement $m\rightarrow m(\psi )$. The field
equations remain the same, except that (\ref{fin13}) will contain terms
proportional to $N_{0}\frac{\partial V}{\partial \psi }$ and $N_{0}\frac{%
\partial m}{\partial \psi }$, and $\left. \frac{\partial \left(
mN_{0}\right) }{\partial x}\right| _{x=z_{a}}=0$ will modify (\ref{eqzeqs}).
While solving the field equations between the particles is straightforward %
\cite{pelzer}, implementation of the matching conditions is not, and so the
existence of an equilibrium solution in the general case remains an
interesting open question.

The solution we have obtained depends quite sensitively on the matching
conditions, and these in turn depend upon the equality of the point masses
and their proper separations. \ Perturbations from this equilbrium case
would be interesting to study, as they would form model inhomogeneous
self-gravitating cosmological systems. \ For sufficiently large positive $%
\Lambda $, we expect that cosmological expansion will overcome the tendency
to gravitationally clump, with the bodies maintaining some form of thermal
equilibrium diluted by the overall expansion. For negative $\Lambda $, there
should be an instability to formation of a 2d black hole \cite{r3,browndan}.
\ 

\vskip0.25cm

This work was supported by the Natural Sciences and Engineering Research
Council of Canada.

\end{document}